# Machine Learning Predictive Analytics for Social Media Enabled Women's Economic Empowerment in Pakistan

Maryam Arif *, Soban Saeed†
*Department of Management Sciences, COMSATS University Islamabad, Pakistan
Email: fa21-bba-095@isbstudent.comsats.edu.pk
†Department of Electrical Engineering, National University of Sciences & Technology (NUST), Islamabad, Pakistan
Email: ssaeed.bee2021mcs@student.nust.edu.pk

*Abstract*—Our study investigates the interplay between young women's empowerment and Pakistan's economic growth, focusing on how social media use enhances their businesses and drives economic advancement. We utilize a mixed-methods research design, integrating both online and offline random sampling, for our survey of 51 respondents. We also utilized existing datasets consisting of both social media usage $n = 1000$ and entrepreneurship $n = 1092$. Our analysis identifies distinct social media engagement patterns via unsupervised learning and applies supervised models for entrepreneurship prediction, with logistic regression outperforming all other algorithms in terms of predictive accuracy and stability. In social media use, the cluster analysis reveals that at K=2, users form tightly packed, well-separated engagement groups. The results indicate that 39.4 percent of respondents believe social media positively impacts the economy by enabling businesses to generate increased revenue. However, only 14 percent of respondents participate in entrepreneurship, highlighting a substantial gap between digital engagement and business adoption. The analysis indicates that daily social media consumption is widespread with YouTube (66.7 percent) and WhatsApp (62.7 percent) being the most frequently used platforms. Key barriers identified are online harassment, limited digital literacy, and cultural constraints in a patriarchal society such as Pakistan. Additionally, 52.9 percent of respondents are unaware of government initiatives supporting women entrepreneurs, indicating limited policy outreach. In contrast, 52 percent express interest in leveraging social media for future entrepreneurial ventures, highlighting significant untapped potential. Overall, the study demonstrates that targeted interventions addressing access, safety, skills, and policy awareness can enable social media to serve as a catalyst for gender-inclusive economic growth and inform policymakers in Pakistan.

*Index Terms*—Social Media, Women Entrepreneurship, Machine Learning, Economic Growth, Digital Empowerment, Dimensionality reduction

## I. INTRODUCTION

### A. Background and Rationale

Entrepreneurship is widely recognized as a key driver of economic growth and development. Circulation of money occurs when goods and services are exchanged and entrepreneurs respond to market demand by producing them. As a result, entrepreneurship addresses challenges such as unemployment and contributes to creating systems that generate jobs and foster socio-economic transformation. Pakistan has one of the largest youth populations globally, with approximately 63% of its citizens aged 15-33 [11]. However, the country faces significant challenges due to a patriarchal society in which men often discourage women from participating in the labor force [25]. This societal structure creates barriers for aspiring young female entrepreneurs who face restrictions on access to technology and mobility. Consequently, their entrepreneurial potential and opportunities are constrained. This paper examines potential solutions to these challenges. The advent of the internet and the expansion of social media have transformed the business landscape, enabling individuals to sell products and services online and across borders. Traditional marketing methods are often more costly for entrepreneurs than cost-effective strategies offered by social media, such as targeted advertising. Social media platforms also facilitate organic growth through daily engagement and content creation. A comprehensive understanding of the impact of social media on young entrepreneurs, policymakers and educators could yield positive outcomes for Pakistan's economy. Enhancing women entrepreneurs' participation and increasing the employment rate are critical economic indicators of overall national growth. This study also investigates two analytical tasks:

1) Behavioral segmentation in social media engagement
2) Prediction of entrepreneurial outcomes among graduate profiles

Clustering and machine learning models are employed to identify patterns and evaluate the contributions of academic background and skills to entrepreneurship. The objective is to develop a transparent pipeline that integrates data analysis with predictive modeling.

## II. LITERATURE REVIEW

The intersection of social media, entrepreneurship, and gender empowerment has garnered increasing attention in recent years, particularly in developing economies like Pakistan, where digital transformation intersects with entrenched socio-cultural norms. Existing research highlights how social media platforms serve as both enablers and disruptors for female entrepreneurs, creating a complex landscape of opportunities and challenges.

Research that explains the positive impact of technology and the internet on the lives of women in patriarchal societies is gaining attention. Some studies, like Social media AS a conduit for women entrepreneurs in Pakistan [29], show social media as a means for women entrepreneurs to bypass traditional hurdles such as lack of mobility and access to markets. This is in tandem with evidence from all over the globe that shows how digital technology eases the market entry challenges to one's own business, especially to economically and socially excluded [28]. In Pakistan, women with home-based businesses use digital technology, especially social media, for product marketing, client networking and transaction management [19].

The availability of digital technologies has transformed economic activities on a global scale. However, Ahmed's 2025 report acknowledges that women's economic activities, for the most part, remain dominated by poorly structured technologies. Limited digital competence, lack of affordable smartphones and internet connection, and negative attitudes regarding women's economic activities are critical barriers that Ahmed et al. [1] identify as key barriers. Such barriers are evident in Pakistan, where women's mobile internet usage is only 26% and 49% for men [23]. The urban-rural divide further exacerbates these disparities, with rural women facing compounded challenges of infrastructure limitations and conservative social norms [18]

Another potential area of study is social media and its algorithms. Z. Ashraf et al. [6] study confirms that women entrepreneurs, particularly in female-dominated businesses such as crafts and home-based food, are economically and socially disadvantaged by social media platform algorithms. Such algorithmic discrimination in its social context of online harassment creates a "hostile digital ecosystem" [2] and further discourages women's participation. The social media and women's entrepreneurship nexus in relation to cultural factors is the most significant of these unexplored social media characteristics.

Research by Duaa Rehman and Urooj Qama [21] highlights how patriarchal norms influence women's digital networking behaviors, with many entrepreneurs relying on male relatives to manage online transactions to maintain social respectability. This finding resonates with broader scholarship on how women in conservative societies navigate technology use within culturally acceptable boundaries [8].

The COVID-19 pandemic marked a significant turning point in this landscape, accelerating digital adoption while also exposing vulnerabilities. Studies such as Women Entrepreneurs and The Usage of Social Media for Business Sustainability in the Time of Covid-19 [20] document how Pakistani women entrepreneurs turned to social media as physical market access became restricted, leading to what [7] describes as "forced digitalization". While this shift created new opportunities, it also intensified challenges like online competition and the need for rapid digital skills acquisition.

Government and institutional responses to these challenges remain uneven. Research like Barriers, Opportunities, and Policy Implications for Women Entrepreneurs in Pakistan [27] critiques existing support programs for failing to address the specific needs of young female digital entrepreneurs, particularly regarding cybersecurity and financial inclusion. This gap is particularly concerning given findings from N. Saif et al. [24] that show how limited digital marketing skills constrain business growth potential.

Theoretical frameworks applied to this phenomenon vary across studies. Some scholars employ feminist political economy approaches to analyze how digital platforms both challenge and reproduce gendered economic inequalities [15]. Others utilize the Capability Approach to assess how social media expands or constrains women's substantive freedoms in the economic sphere [13]. These diverse theoretical lenses reflect the multidimensional nature of the issue, encompassing technological, economic, and socio-cultural dimensions.

Methodologically, existing research on this topic tends to fall into two categories:

1) Quantitative studies measuring adoption rates and business outcomes (e.g, [10])
2) Qualitative explorations of lived experiences (e.g, [13])

While both approaches yield valuable insights, there remains a need for more mixed-methods research that can capture both statistical patterns and contextual nuances, a gap this study seeks to address.

Our research builds on and extends this existing literature in varying ways. While previous research has explored the opportunities and challenges of social media for women entrepreneurs in isolation, this study examines the interaction of both components. Addressing social media usage in business is particularly relevant for young entrepreneurs, yet there is a gap in research for young female entrepreneurs (14-25) who are digital natives and are shaping the future of entrepreneurship in Pakistan. Using a mixed-methods approach, we were able to add the generalization of surveys to the richness of qualitative data collected, which has not been done in previous studies. With this, we have a better understanding of the phenomenon than previous studies which utilized a single research method.

The precedent research has determined that principal component analysis is an effective technique to use in order to extract salient features of high dimensional datasets [12], systematically proportion of the K-Means clustering technique to calculate user subsets of a dataset by categorizing them into large groups of similar users [14]. The research on entrepreneurship prediction generally utilizes logistic regression, decision tree-based algorithms and ensemble learning based models to model scholarly and cognitive anticipative of entrepreneurial achievements [26], [9]. Predictive pipelines in the educational and behavioral domains have been improved with artificial intelligence techniques, including SHAP. This paper combines PCA-based clustering and SHAP-enhanced predictive models on two different datasets [17].

III. METHODOLOGY

This study employed a quantitative methodology to examine the relationship between social media use and entrepreneurial activities among young female entrepreneurs in Pakistan. The data collection strategy for the quantitative methodology. Both online and offline surveys were given to the respondents. The survey contains information essential to our understanding of the study on the social media impact on women entrepreneurs, including social media use, its effects on young women and its impact on economic growth.

### A. Sampling Strategy

This study includes 51 young female respondents aged 14 to 25 from across Pakistan, selected through a quantitative survey. The sample consists of both active entrepreneurs and social media users with entrepreneurial aspirations. Findings indicate that this age range encompasses critical developmental stages for acquiring digital literacy. The sample size was determined through pragmatic considerations of resource constraints and the need for in-depth qualitative analysis [22].

*1) Geographic distribution:* The sample includes both urban and semi-urban locations in Islamabad, Rawalpindi, Faisalabad, and Sargodha, thereby capturing diverse socio-economic contexts. Random sampling was conducted through women's entrepreneurship networks and educational institutions to ensure representation across various industry sectors, including but not limited to e-commerce (23%), fashion (19%), food services (15%), and technology (8%).

*2) Data Collection Instruments:* The survey instrument was across four domains:

- **Demographic characteristics**: Age, education level, geographic location
- **Social media usage patterns**: Platform preferences, frequency, duration
- **Entrepreneurial engagement**: Business type, years of operation, revenue streams
- **Perceived impacts**: Economic benefits, challenges, confidence levels

Attitudes were measured using Likert-scale questions (1 = strongly disagree to 5 = strongly agree), while open-ended questions collected qualitative insights. The survey was administered in English and Urdu through both Google Forms and paper questionnaires to accommodate varying literacy levels.

*3) Ethical Considerations:* Ethical considerations were addressed in this study. Participation in the survey was voluntary, and all participants completed the survey with a clear understanding of the study's purpose. The survey procedures ensured the confidentiality and anonymity of participants' responses.

### B. Dataset Description

The first dataset consists of 1,000 social media users and includes features such as daily minutes spent, number of posts, likes, follows, and application category. This dataset is publicly available on Kaggle [16]. The second dataset comprises 1,092 graduate profiles, each containing academic indicators such as GPA, SAT scores, university ranking, certifications, skill scores, employment outcomes and a binary label for entrepreneurship.

The flowchart below demonstrates the complete pipeline, ranging from the Dataset to clustering and predictions.

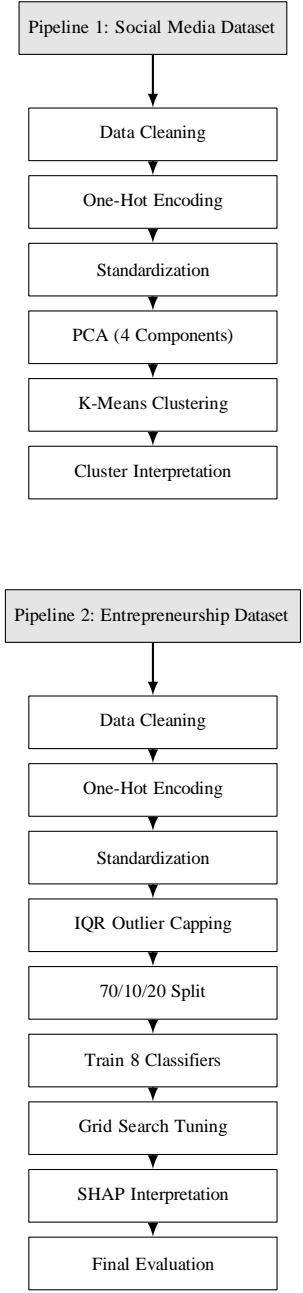

Fig. 1. Compact dual-pipeline methodology illustrating preprocessing and PCA-based clustering for the social media dataset (top), and the supervised learning pipeline for the entrepreneurship dataset (bottom).

### C. Data Preprocessing

Both datasets were examined for missing values and duplicate records. Categorical attributes were encoded using a one-hot encoding scheme, while numerical variables were standardized according to,

$$\tilde{x}_{ij} = \frac{x_{ij} - \mu_j}{\sigma_j}, \quad (1)$$

where $\mu_j$ and $\sigma_j$ denote the mean and standard deviation of feature $j$, respectively.

Potential outliers in the numerical features were treated using Interquartile Range (IQR)–based capping. The en-trepreneurship dataset was partitioned into training (70%), validation (10%), and testing (20%) subsets.

### D. Dimensionality Reduction (PCA) for Social Media Dataset

To identify the interaction structure, Principal Component Analysis (PCA) was applied to the standardized social media dataset. Let $X \in \mathbb{R}^{n \times p}$ denote the centered feature matrix, and let

$$C = \frac{1}{n-1} X^\top X \quad (2)$$

represent its covariance matrix. PCA computes eigen pairs by solving

$$C v_k = \lambda_k v_k, \quad (3)$$

where $\lambda_k$ and $v_k$ are the $k$th eigenvalue and eigenvector, respectively.

The projection onto the first $K$ components is given by

$$Z = X V_K, \quad (4)$$

where $V_K = [v_1, \ldots, v_K]$.

Four components explaining approximately 84% of the total variance were retained. As shown in Figure 2, PCA reduces dimensionality by extracting principal components.

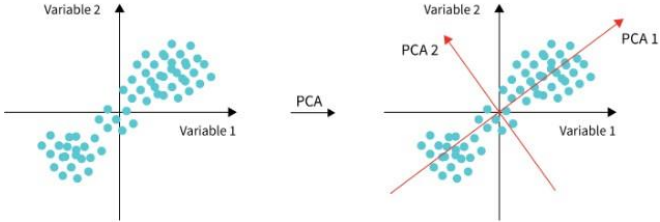

Fig. 2. Conceptual illustration of Principal Component Analysis (PCA). Source: Scaler

### E. Clustering over Dimensionally reduced Data

The PCA representation $Z$ was clustered using the K-Means algorithm, which solves the optimization problem

$$\min_{\mu_1, \ldots, \mu_K} \sum_{i=1} \min_j \| z_i - \mu_j \|^2, \quad (5)$$

where $\mu_j$ denotes the centroid of cluster $j$. Silhouette scores and inertia curves consistently indicated $K = 2$, revealing two distinct groups: a low-interaction user cluster and a high-engagement user cluster.

Five classifiers were trained for the predictive modeling stage, including Logistic Regression, Decision Tree, Random Forest, $k$-Nearest Neighbors (kNN), and Support Vector Machines (SVM). Grid search was applied to extract the best parameters for each model.

## IV. MODEL SELECTION

The following models were tested for entrepreneurial predicton:

*1) Logistic Regression:* Entrepreneurship probability is modeled as

$$P(y = 1 \mid x) = \sigma(w^\top x + b), \qquad \sigma(t) = \frac{1}{1 + e^{-t}}, \quad (6)$$

with an L2-regularized loss function defined by

$$\mathcal{L}(w) = -\sum_i [y_i \log \sigma_i + (1 - y_i) \log(1 - \sigma_i)] + \lambda \|w\|_2^2. \quad (7)$$

Table IV presents the complete hyperparameter grid used for Logistic Regression.

| Parameter | Values |
|---|---|
| Penalty | {l1, l2} |
| C | {0.001, 0.01, 0.1, 1, 10} |
| Solver | liblinear |
| Max Iterations | 2000 |
| Scoring Metric | f1 |
| Cross-Validation Folds | 5 |
| n_jobs | −1 |
| Random State | 42 |

TABLE I
LOGISTIC REGRESSION GRID SEARCH PARAMETERS

*2) Decision Tree:* The model recursively partitions the feature space by minimizing the Gini impurity measure

$$G(t) = 1 - \sum_c p(c \mid t)^2, \quad (8)$$

where $p(c \mid t)$ denotes the proportion of class $c$ in node $t$. Table IV summarizes the grid used for Decision Tree tuning.

| Parameter | Values |
|---|---|
| Criterion | {gini, entropy, log_loss} |
| Max Depth | {None, 5, 10, 20, 50} |
| Min Samples Split | {2, 5, 10, 20} |
| Min Samples Leaf | {1, 2, 5, 10} |
| Max Features | {None, sqrt, log2} |
| Max Leaf Nodes | {None, 10, 50, 100} |
| Min Impurity Decrease | {0.0, 0.01, 0.1} |
| Splitter | {best, random} |
| Class Weight | {None, balanced} |
| CCP Alpha | {0.0, 0.01, 0.1} |

TABLE II
DECISION TREE GRID SEARCH PARAMETERS

*3) k-Nearest Neighbors:* The kNN classifier assigns a label according to the majority class among the $k$ nearest samples measured under the Euclidean distance metric.

All models were tuned using grid-search hyperparameter optimization. Additionally, SHAP (SHapley Additive exPlanations) values were computed for all classifiers to quantify feature contributions and interpret model behavior.

Table IV provides the hyperparameter variations evaluated for kNN.

| Parameter | Values |
|---|---|
| n_neighbors | {3, 5, 7, 9} |
| Weights | {uniform, distance} |
| Metric | {euclidean, manhattan} |

TABLE III
$k$-NEAREST NEIGHBORS GRID SEARCH PARAMETERS

*4) Random Forest:* Random Forest constructs an ensemble of $M$ decision trees, each trained on a bootstrapped sample of the data. The final prediction is obtained via majority voting:

$$\hat{y} = \text{mode}\{h_m(x)\}_{m=1}^{M}. \tag{9}$$

Table IV lists the full hyperparameter search space for Random Forest.

| Parameter | Values |
| --- | --- |
| n estimators | {50, 100} |
| Max Depth | {5, 10, None} |
| Min Samples Split | {2, 5} |
| Min Samples Leaf | {1, 2} |
| Max Features | {sqrt, log2, None} |
| Bootstrap | {True, False} |
| Criterion | {gini, entropy} |
| Class Weight | {None, balanced} |
| Random State | 21 |

TABLE IV
RANDOM FOREST GRID SEARCH PARAMETERS

*1) Support Vector Machines:* Table IV-1 shows the grid used for Support Vector Machine optimization.

| Parameter | Values |
| --- | --- |
| C | {0.1, 1, 10} |
| Gamma | {scale, auto} |
| Kernel | rbf |
| Probability | False |

TABLE V
SUPPORT VECTOR MACHINE GRID SEARCH PARAMETERS

## V. DATA ANALYSIS

The quantitative data collected from survey responses is analyzed to evaluate several dimensions, including economic growth, social media influence, business performance, innovation and adaptability, networking and collaboration, skill development, challenges and obstacles and long-term sustainability.

| Variable | Categories | &age |
| --- | --- | --- |
| Age Group | 14–18 | 32.6 |
| | 19–24 | 54.9 |
| | 25 | 12.5 |
| Education Level | Secondary School | 41.2 |
| | University | 58.8 |
| Entrepreneurial Status | Engaged in Entrepreneurship | 14.0 |
| | Not Engaged in Entrepreneurship | 86.0 |

TABLE VI
DEMOGRAPHIC PROFILE OF SURVEY RESPONDENTS SHOWING AGE GROUP DISTRIBUTION, EDUCATION LEVEL, AND ENTREPRENEURIAL STATUS.

## VI. KEY FINDINGS

### A. Economic Growth

The findings revealed a wide range of perspectives, as data were collected from multiple regions of Pakistan. A substantial proportion of respondents expressed confidence in the positive role of social media in promoting economic growth. However, 20.8% of participants were uncertain about its impact, reflecting ambiguity. This uncertainty may result from factors such as limited confidence in social media's effectiveness, insufficient proficiency in its use, concerns about Pakistan's future economy, or a lack of exposure to successful social media-driven business models. Approximately 21% of respondents believed that social media has no significant effect on the economic potential of young female entrepreneurs. This perspective may stem from limited familiarity with social media or a preference for traditional business methods, which are perceived as more effective. Additionally, 18.8% of respondents indicated that social media negatively affects Pakistan's economic growth prospects. This negative perception is likely rooted in concerns about online harassment, information overload, and the prevalence of unverified content, all of which may impede entrepreneurial progress. These findings suggest that some respondents may lack familiarity with social media or prefer traditional business methods, which they consider more effective. To gain a comprehensive understanding, it is important to consider the perspectives of both entrepreneurial and non-entrepreneurial respondents. Including non-entrepreneurs in the survey broadens the range of perspectives and clarifies the diverse perceptions of social media's influence on entrepreneurship in Pakistan. While a notable portion of young female entrepreneurs view social media positively for economic growth, significant numbers also hold negative or uncertain views.

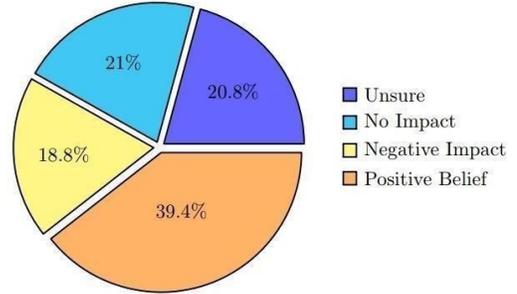

Figure 1: Perceptions on Social Media Impact

### B. Social media usage

Social media has become a significant component of contemporary life. The collected data reveal patterns in the frequency and duration of respondents' social media use. Approximately 76.5% of participants report daily social media use, indicating its central role in their routines and activities. These individuals regularly consume content, build connections, and interact with others. An additional 13.7% of respondents use social media several times a week, suggesting that, while not daily users, they still consider these platforms important in their lives. Only 4.8% report using social media once a week, and about 2% use it rarely. These findings demonstrate that the majority of individuals engage with social media daily, while a small proportion use it less frequently. The data highlight the pervasive integration of social media into daily life, particularly among students and young people. This widespread usage underscores the potential of social media as a positive tool for the growth of female entrepreneurs' businesses.

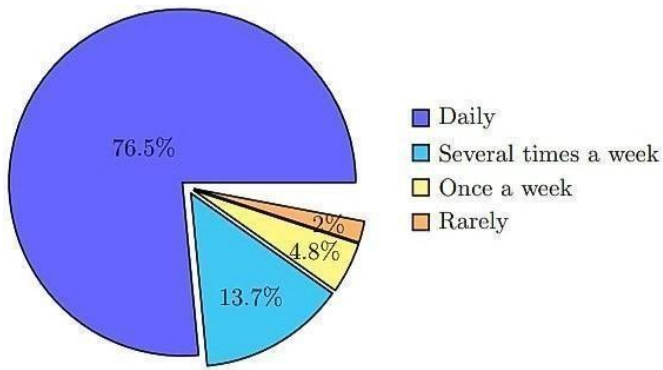

Figure 1: Frequency of Social Media Usage

## C. Number of Hours Spent on Social Media

Table 1: Number of Hours Spent on Social Media

| Time Spent | Percentage |
|---|---|
| 1-2 hours daily | 39.2% |
| Less than 1 hour daily | 15.7% |
| 2-4 hours daily | 15.7% |
| More than 4 hours daily | 29.4% |

Table 2: Entrepreneurial Activities

| Engagement in Entrepreneurial Activities | Percentage |
|---|---|
| No | 86% |
| Yes | 14% |

Table 3: Perception of Social Media's Influence

| Confidence Level | Percentage |
|---|---|
| Very confident | 17.6% |
| Somewhat confident | 17.6% |
| Neutral | 51% |
| Not very confident | 3.9% |
| Not confident at all | 9.8% |

According to the survey, approximately 39.2% of respondents use social media for 1-2 hours daily. A further 15.7% use it for less than 1 hour per day, indicating relatively low engagement, while another 15.7% report using social media for 2-4 hours daily. About 29.4% of respondents use social media for more than 4 hours each day. These findings suggest that social media usage is prevalent among women and teenagers, highlighting its potential as a marketing tool for Pakistani women. Regarding entrepreneurial activities, approximately 86% of teenage and young women reported not engaging in them, whereas 14% reported participating in them in Pakistan. When asked about their confidence in leveraging social media's impact, about 51% of women expressed a neutral stance, reflecting uncertainty among teenagers and young women about utilizing social media for business purposes. In contrast, 17.6% were somewhat confident, and an equal proportion, 17.6%, were very confident and expressed positive attitudes toward using social media to promote their entrepreneurial ventures. Additionally, 9.8% reported no confidence at all, which may be attributed to negative past experiences or skepticism regarding the effectiveness and benefits of social media for their current or future businesses.

## D. Government initiatives

In response to this opportunity, the government of Pakistan has implemented several initiatives and programs to help young women entrepreneurs use social media for business development. Notable examples include SheMeansBusiness, Womenx, and the Women Entrepreneurship Development System (WEDS). Despite these efforts, 52.9% of respondents reported a lack of awareness of such programs, while 47.1% indicated awareness of government initiatives. Furthermore, when asked about the use of social media platforms for product or service promotion, 72.5% responded negatively, while 27.5% affirmed their use.

## E. Innovation and Adaptability

To assess the impact of social media on gender equality in entrepreneurship, participants were asked to respond with either "Yes" or "No" regarding their beliefs. Among respondents, 24.5% indicated that social media does not provide equal opportunities for male and female entrepreneurs, while 75.5% believed it does. Qualitative analysis was also conducted. The survey results suggest that many female entrepreneurs experience gender bias on social media, as evidenced by lower follower counts and lower engagement rates. Female participants reported experiencing online harassment, which negatively affects self-esteem and may contribute to mental health challenges. Maintaining a work-life balance while managing a business on social media is particularly challenging for women, given the traditional mindset of Pakistani society and the gender roles imposed on women. Several respondents also noted that limited access to funding serves as a barrier to growth on social media. While partnerships and collaborations can facilitate networking and engagement with like-minded individuals, the lack of such opportunities often leaves female entrepreneurs feeling isolated and unsupported. Consequently, insufficient networking opportunities constitute a significant barrier.

## F. Challenges in Leveraging Social Media for Entrepreneurial Pursuit

Among the respondents, 18.8% reported challenges in leveraging social media for their entrepreneurial pursuits. While the majority (60.4%) said they do not use social media for their entrepreneurial activities. 20.8% said they did not use social media for their entrepreneurial activities. Harassment on social media is one of the leading issues for women entrepreneurs. Harassment takes many forms, such as online abuse, hate speech, cyberbullying, etc., which can have many consequences on their mental health.

### G. Consequences of Harassment

Reputation damage resulting from false allegations can significantly harm both the brand image and customer trust associated with female entrepreneurs. Research indicates that women are more likely than men to experience emotional distress, including stress and depression, as a result of harassment. Such experiences can adversely affect the health of women entrepreneurs. Concerns for personal safety may intensify if harassment escalates, potentially leading to physical threats. Another prevalent challenge is the lack of resources and limited access to technology, which restricts the effective use of social media for business development. Financial constraints, slow internet speeds, and limited access to essential equipment and software further hinder the ability to provide timely, high-quality customer service. These barriers not only place women entrepreneurs at a competitive disadvantage but also make it difficult to afford essential technological tools. Data suggest that cultural and social norms present additional obstacles when entrepreneurs utilize social media. However, the specific nature of these challenges remains unclear, as the data only references general issues without detailed descriptions. Cultural differences may complicate entry into global markets, necessitating tailored content and language strategies to engage a diverse, international audience. Traditional gender roles persist, particularly in Pakistan, where only 1% of women participate in business, reflecting a significant gender gap and entrenched societal attitudes.

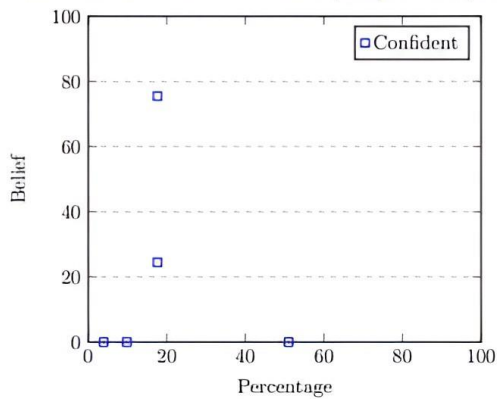

Figure 1: Social Media's Influence on Gender Equality in Entrepreneurship

Privacy concerns are also prominent among female entrepreneurs, with survey results indicating a lack of confidence in the safety of their personal information. As globalization increases, language barriers become more pronounced when entering new markets. Therefore, women entrepreneurs must adapt their social media strategies to address the cultural and linguistic needs of their target audiences.

### H. Willingness to Use Social Media for Future Product Sales

Among the respondents, 52% indicated that they would consider selling their products through social media in the future, suggesting a recognition of its potential business advantages. Approximately 36% of respondents expressed uncertainty or reservations about using social media for their own purposes, possibly due to a lack of confidence or limited exposure to successful examples. In contrast, 12% reported no intention of utilizing social media for future product sales.

*1) Growing Importance of Social Media (43.1%):* About 43.1% expressed a strong belief in the importance of social media for young women entrepreneurs in Pakistan. That is about 51 out of 21 respondents. After all, social media has become an essential tool that is to be used for a broader audience.

### I. Building Brand Awareness

The respondents recognized the significance of brand visibility in the current competitive environment. Social media offers young women entrepreneurs opportunities to establish their brands and provides a platform to share their narratives with fewer obstacles or financial limitations, challenges that are more pronounced for women operating offline in Pakistan. The global expansion of social media has increased audience reach and facilitated connections with diverse communities. As a result, women entrepreneurs can now market and sell their products internationally.

### J. Effective Customer Engagement

Social media offers entrepreneurs the opportunity to connect with their consumers and build meaningful relationships with their audiences. This includes benefits such as making it easier to address customer queries and enhancing the brand's level from the customer's perspective.

### K. Powerful Marketing Tool

The respondents regard social media as a versatile marketing tool of increasing importance. They recognize the advertising opportunities and exposure social media provides, as well as its capacity to facilitate targeted marketing campaigns and yield higher returns on investment.

### L. Showcasing Products or Services

Younger women entrepreneurs believe that social media is an ideal platform for showcasing their products to the world. They can use unique selling points and, on occasion, marketing that can generate more sales.

### M. Social Media Insignificance

However, 5 respondents (9.8%) believe that social media cannot play a significant role in the success of young women entrepreneurs. There could be multiple reasons for this conclusion, most people consider traditional businesses real businesses, and these business practices hold more value in the respondents' view. Thus, social media does not align with their success. Some women entrepreneurs consider social media a distraction. Therefore, they would not like to spend a lot of time figuring things out themselves, as it can be time-consuming. Pakistan's internet penetration rate stood at 36.5%, indicating that most aspiring women entrepreneurs lack internet access. Thus, they will not be able to use social media tools and techniques effectively. About 47.1% of respondents selected "maybe" as their response regarding the impact and success of social media on young women entrepreneurs. Thus, uncertainty may be from various factors.

*N. Lack of knowledge about effective strategies*

Several respondents indicated limited knowledge or understanding regarding the effective use of social media to promote business growth.

*O. Rapidly changing landscape of Digital marketing*

The landscape of social media and digital marketing is continually evolving, creating an increasingly competitive environment. This dynamic creates uncertainty among respondents about the long-term impact of social media.

Respondents identified both advantages and disadvantages associated with using social media for business purposes. While social media serves as a powerful tool, it also presents certain challenges. Analysis of the survey data provides detailed insights into the use of specific platforms. Facebook remains popular, with 19 out of 51 respondents (approximately 37.3%) reporting frequent use.

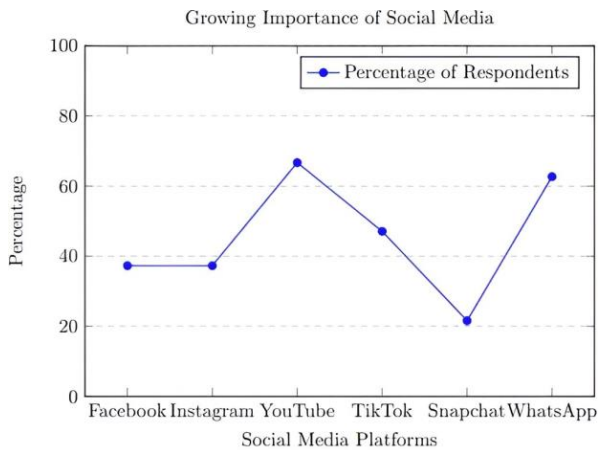

Figure 1: Growing Importance of Social Media

Despite the emergence of new applications, Facebook maintains its position due to its diverse user base and features such as photo sharing and connectivity for staying informed about news and events, as well as for sharing experiences. This functionality is particularly significant for female entrepreneurs, as it facilitates audience engagement through experience sharing.

Instagram was also frequently used by 19 respondents, representing approximately 37.3% of the sample. It is among the most widely used applications, particularly among younger audiences. Features such as Reels and Stories contribute to high levels of user engagement. YouTube is widely used, with 66.7% of respondents reporting use (approximately 34 out of 51 individuals). The platform offers a diverse range of content, including vlogs, educational materials, and entertainment, making it one of the most frequently used applications among younger users. TikTok, a relatively new social media platform, attracted 47.1% of respondents (21 out of 51), reflecting its popularity among younger demographics and its rapid growth in recent years. The emergence of new content creators has contributed to the platform's expanding diversity of content.

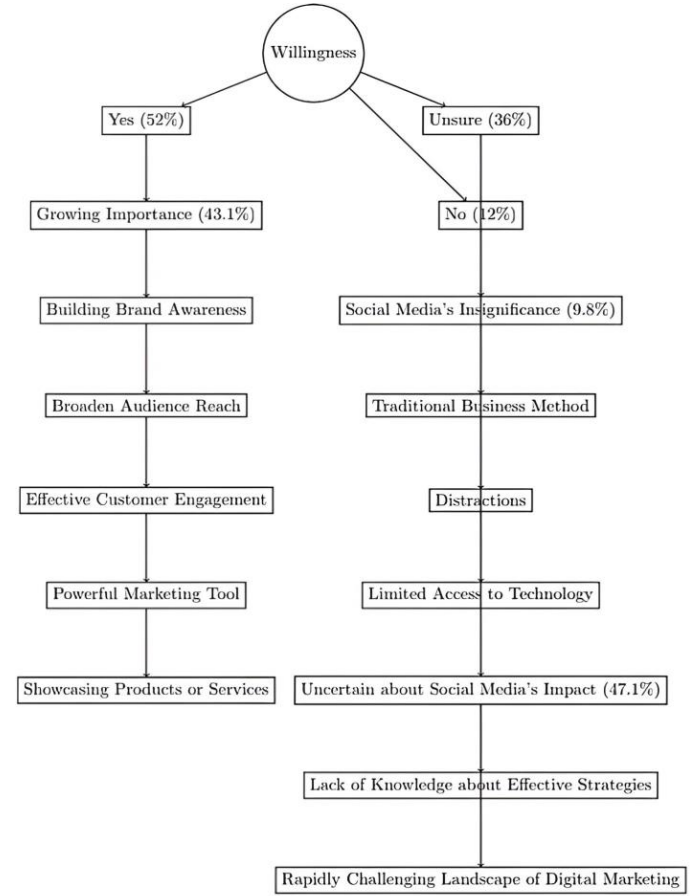

Figure 1: Willingness to Use Social Media for Future Product Sales

Snapchat was used by approximately 21.6% of respondents (11 out of 51), a trend attributed to its appeal to younger demographics and the engaging nature of Snap Stories. WhatsApp was reported as a frequently used application by 62.7% of respondents (32 out of 51). It is one of the most popular communication tools in Pakistan, particularly among younger users, thanks to free texting. Businesses also recognize WhatsApp as a valuable tool for customer support and engagement. The use of social media varies according to factors such as age, cultural background, and personal preferences. As the social media landscape continues to evolve, user preferences are constantly changing. Respondents noted the integration of social media into various aspects of their lives, highlighting its roles beyond personal use, including political and business applications. Given the widespread adoption of social media, women in developing countries, in particular, may benefit from its opportunities.

## VII. RESULTS

*A. Machine Learning on Social Media Usage Dataset*

The pipeline generated a systematic evaluation of user behaviour using 100 data records. The analysis included de-scriptive statistics, dimensionality reduction, and unsupervised learning techniques.

*1) Statistical Analysis:* The statistics showed a strong variation in behavior particularly in Daily minutes spend and likes per day. These variation serves as the base for clustering. The complete summarization of dataset is depicted in table VII.

| METRIC | COUNT | MEAN | STD | MIN | 25% | 50% | 75% | MAX |
|---|---|---|---|---|---|---|---|---|
| DAILY MIN | 1000 | 247.36 | 146.37 | 5 | 112.75 | 246 | 380.5 | 500 |
| POST | 1000 | 10.27 | 6.12 | 0 | 5 | 10 | 16 | 20 |
| LIKES | 1000 | 94.68 | 57.56 | 0 | 44.75 | 94 | 142 | 200 |
| FOLLOWS | 1000 | 24.69 | 14.84 | 0 | 12 | 24 | 38 | 50 |

TABLE VII
STATISTICAL SUMMARY OF USER ACTIVITY METRICS

Fig 3 demonstrate the correlation among features in the social media usage dataset. The plot shows strong non-correlation among features.

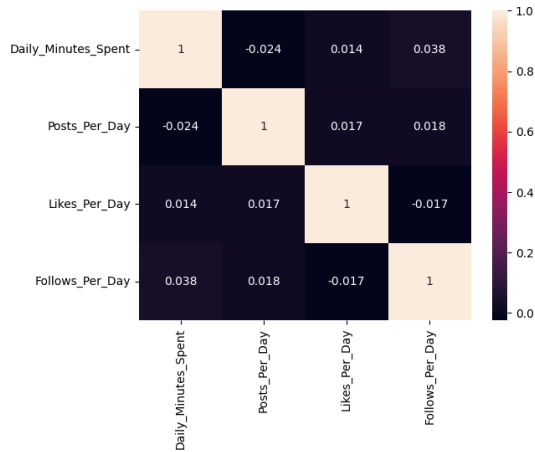

Fig. 3. Heatmap to demonstrate correlation among features

*2) PCA Dimensionality Reduction:* The principal component analysis extracted four components that explain 84.4% of the variance. The PC1 captured heavy loadings from Daily Minutes and Follows per day, PC2 exhibited high contributions from Posts per day and Likes per day.

The scree plot 4 shows that variance stabilizes after the fourth principal component, indicating that key patterns in social media behavior are effectively summarized in a reduced feature space.

The four-dimensional PCA visualization illustrates clear groupings of users, suggesting the presence of meaningful and consistent engagement patterns within the social media dataset.

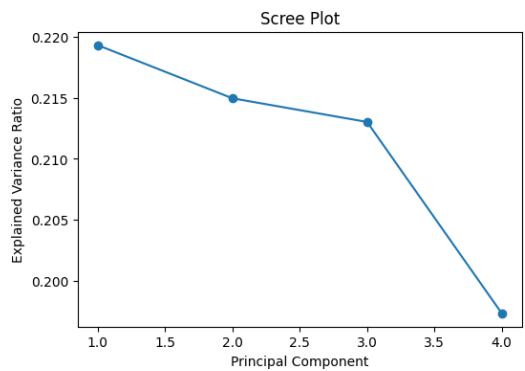

Fig. 4. Scree Plot for Principal Components

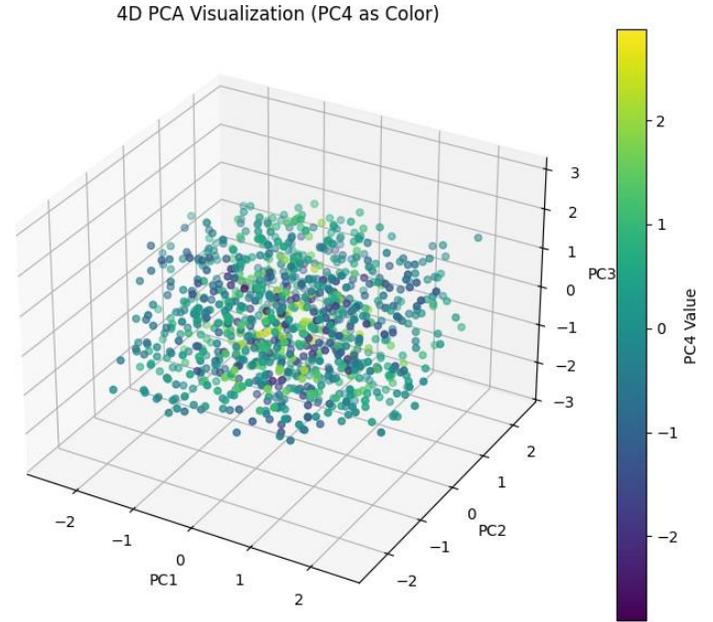

Fig. 5. 4D plot to visualize Principal Component Analysis

A loading matrix summarizing variable contributions is shown in Fig 6. PC1 represents engagement intensity, PC2 and PC3 correspond to posting and liking behaviors, and PC4 indicates mixed engagement. Platform effects are minimal.

Fig. 6. Loading obtained from Principal Component Analysis

*3) Cluster Selection:* The optimal value of clusters (K) for K-Means was determined using Elbow Method and Silhouette Scores. The Silhouette scores indicated the optimal performance and maximum separation at K = 2 as shown in fig 7.

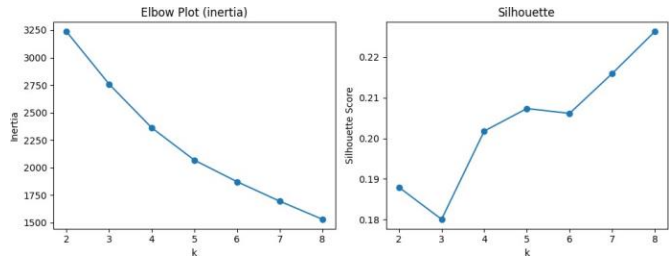

Fig. 7. Elbow plot and Silhouette Scores indicating cluster value.

*4) Cluster Characterization:* The clustering produced two groups, which are summarized in 8. These clusters differ in engagement levels and rate of social actions,

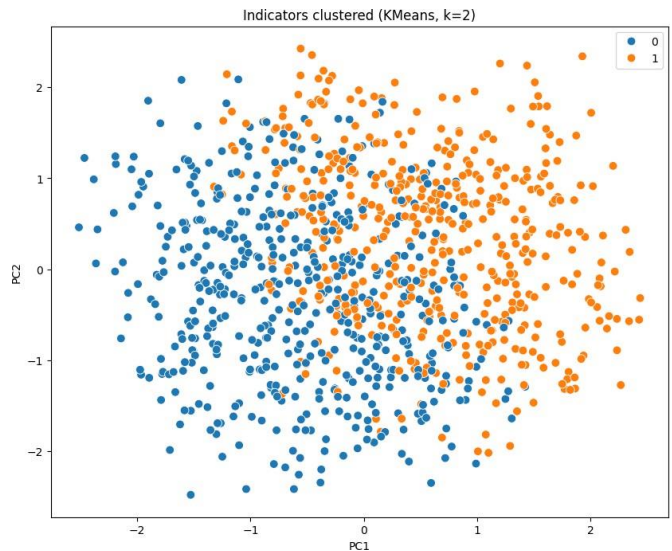

Fig. 8. Clusters at K = 2

*5) Interpretation of Clusters:* Cluster 0 demonstrated moderate posting and liking behavior but low following activity. This cluster showed a content viewer who interacts but never tends to expand their network.

Cluster 1 displayed high following activity and increased socialization. These people represent active network builders.

### B. Machine Learning on Entrepreneurship Dataset

The evaluation examined the prediction efficiency of five models trained on the entrepreneurship dataset. The complete pipeline comprises statistical analysis, encoding, standardization.

*1) Statistical Analysis:* Statistical Analysis revealed balanced class proportions for the target variable (Entrepreneurship). This class balance is showcased in figure 9.

Moderate right skew is reported in the Starting Salary feature within the dataset. The skew is demonstrated in the table VII-B1 and visualized through histograms in fig 11, While the rest of the numerical features showed limited multicollinearity and normal distributions. This non-correlation is reported in fig 10.

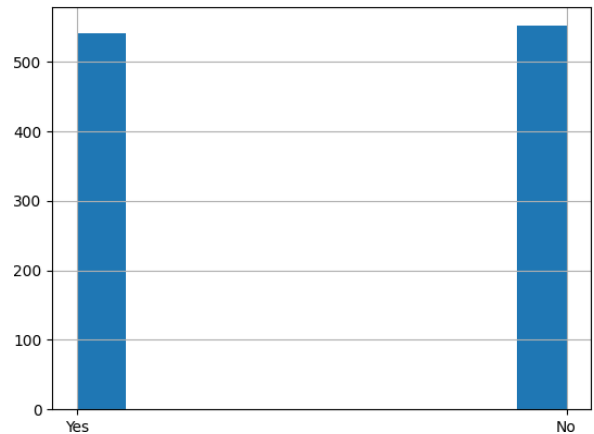

Fig. 9. Histogram representing class balance

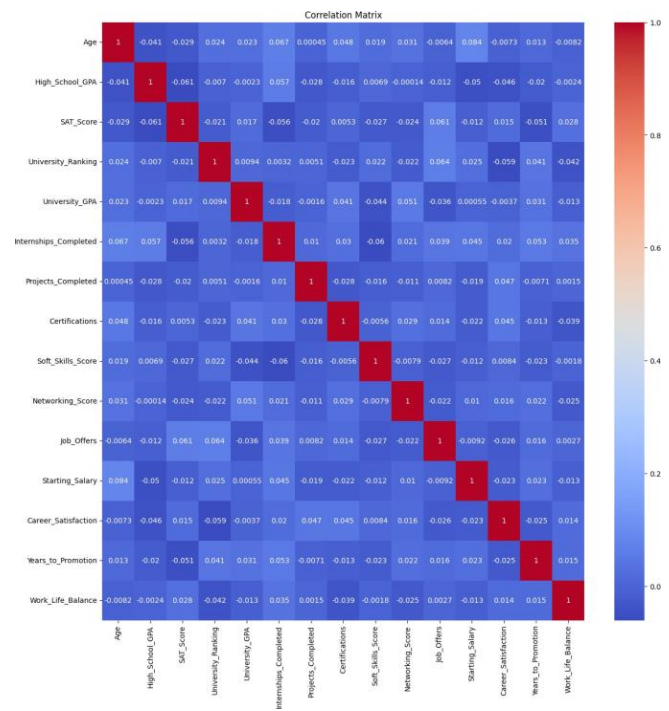

Fig. 10. Heatmap demonstrating the correlation among features

| Feature | Skewness | Distribution Category |
|---|---|---|
| Age | 0.051 | Approximately Normal |
| High School GPA | -0.047 | Approximately Normal |
| SAT Score | -0.019 | Approximately Normal |
| University Ranking | -0.019 | Approximately Normal |
| University GPA | -0.109 | Approximately Normal |
| Internships Completed | 0.022 | Approximately Normal |
| Projects Completed | 0.013 | Approximately Normal |
| Certifications | 0.031 | Approximately Normal |
| Soft Skills Score | 0.028 | Approximately Normal |
| Networking Score | -0.053 | Approximately Normal |
| Job Offers | 0.026 | Approximately Normal |
| Starting Salary | 0.703 | Moderate Skew |
| Career Satisfaction | -0.022 | Approximately Normal |

TABLE VIII
SKEWNESS ANALYSIS OF ENTREPRENEURSHIP DATASET FEATURES

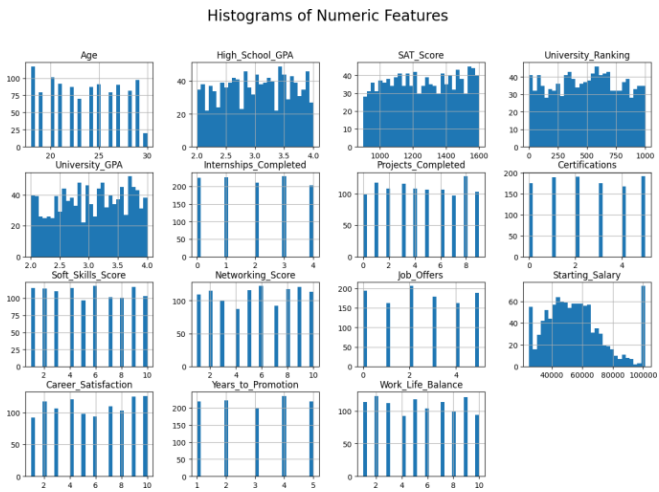

Fig. 11. Histogram of features in Entrepreneurship Dataset

*2) Data Splitting:* The dataset was divided into training, validation, and testing subsets to support model development and evaluation. A total of 764 samples were used for training, 65 for validation, and 263 for testing, each with 25 input features. This split improved accuracy of predictions and the reliability of the model.

*3) Logistic Regression Performance:* Logistic Regression produced the most stable and best results across training, validation and testing. The optimal parameters extracted during Grid Search were L2 regularization and C = 0.001. This model achieved an accuracy of 84.6%, with strong recall for entrepreneurial candidates.

The accuracy curves (fig 12) represent the insights into how model performance varies across evaluated hyperparameter configurations, with both training and validation accuracies. The figure represents stabilization, which indicates that optimal parameter settings were identified, yielding consistent generalization performance without excessive sensitivity to configuration changes.

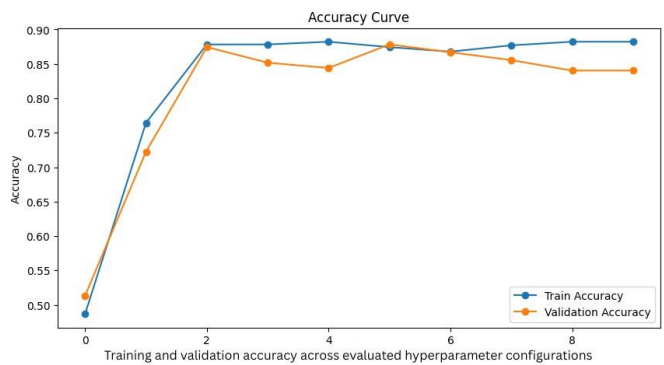

Fig. 12. Training and validation accuracy across evaluated hyperparameter configurations, illustrating performance variation under different model settings.

The corresponding loss curves (figure 13) show a general decline followed by convergence between configurations, thus reflecting an effective optimization under selected parameter ranges. The close alignment between training and validation loss suggests controlled learning behavior and limited overfitting during model selection.

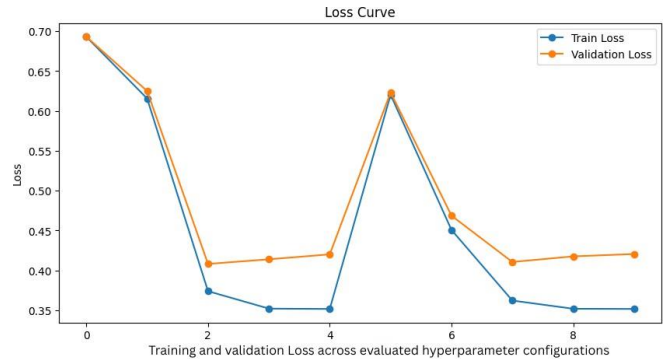

Fig. 13. Training and validation accuracy loss evaluated hyperparameter configurations

The confusion matrix (figure 14) demonstrated a strong classification performance over the entrepreneurship dataset. Only a very small number of misclassifications are reported in the testing subset. Additional evaluation metrics were derived from the confusion matrix to ensure balanced performance across classes.

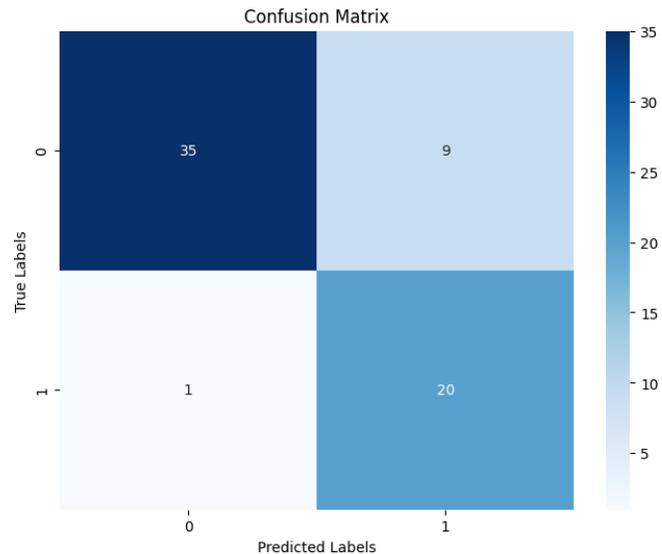

Fig. 14. Confusion matrix reporting performance over the testing subset

The SHAP summary plot (figure 15) represents a small subset of features that dominate the model's predictions with Features 8, 9, and 6 influencing the entrepreneurial prediction. Higher values of these features consistently contribute positively to the predicted likelihood, while lower values suppress it. While the aggregated contribution of remaining features is minimal, it confirms that model decisions are driven by a compact and interpretable feature set rather than diffuse noise.

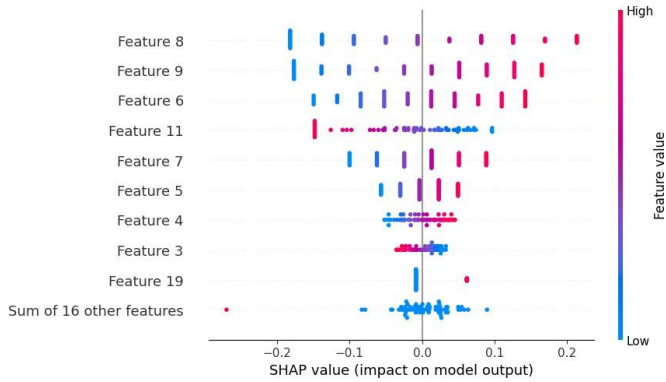

Fig. 15. SHAP plot representing model driving features

*4) Decision Tree Classifier:* The decision tree model demonstrated moderate generalization. Although, training accuracy was high, the validation accuracy reached 61.5%, while 61.5% for testing, representing poor performance and high overfitting under strong regularization setting.

The confusion matrix (figure 16) reported high prediction accuracy for positive class, but poor for the negative class. This indicates that the proposed model correctly classified 23 instances of the non-entrepreneurial class and 17 instances of the entrepreneurial class. The model has a higher tendency to falsely predict entrepreneurship (21 cases) compared to missed entrepreneurial instances (4 cases).

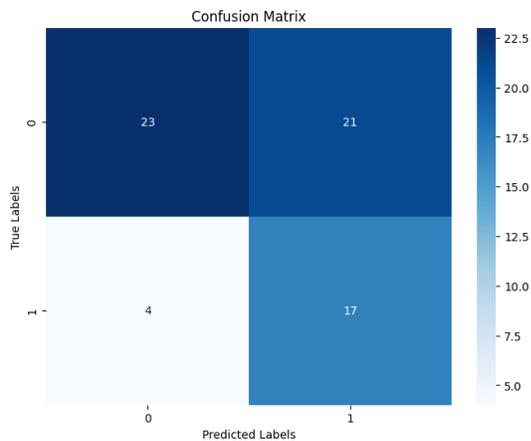

Fig. 16. Confusion Matrix for Decision Tree

*5) Random Forest Classifier:* Random Forest achieved better performance over it's best configuration of 50 trees, depth 5 and log 2 as features, obtained after Grid Search. This model showed 81.5% as testing accuracy with balanced performance across both classes.

The confusion matrix (figure 17) reported high prediction accuracy for positive class, but poor for the negative class. This indicates that the proposed model correctly classified 34 instances of the non-entrepreneurial class and 19 instances of the entrepreneurial class. The model has a higher tendency to falsely predict entrepreneurship (10 cases) compared to missed entrepreneurial instances (2 cases).

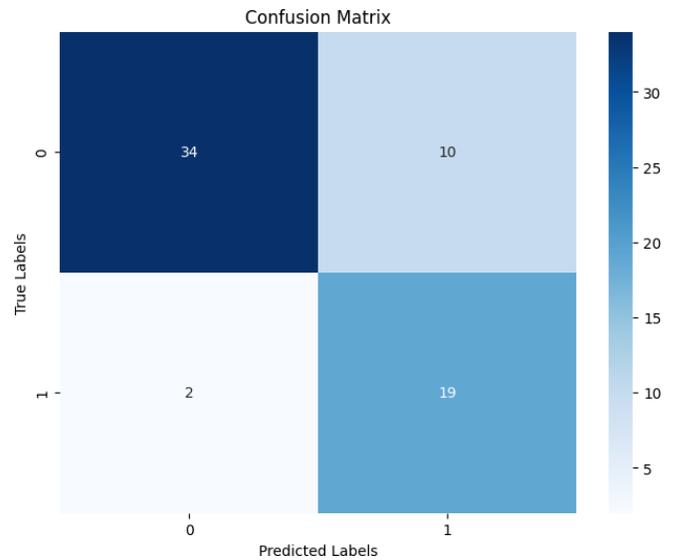

Fig. 17. Confusion Matrix for Random Forest Classifier

*6) K-Nearest Neighbors:* The KNN classifier achieved a notable testing accuracy of 72.3% at its best trained parameter, K = 5. This model has improved recall for entrepreneurial talent, but notable drop in specificity.

The confusion matrix (Fig. 18) indicates that the KNN classifier exhibits the weakest overall performance among the evaluated models. While the model demonstrates comparatively better predictive capability for the positive (entrepreneurial) class, its performance on the negative (non-entrepreneurial) class is notably poor. Specifically, the classifier correctly identified 30 non-entrepreneurial instances and 17 entrepreneurial instances. However, it shows a pronounced tendency toward false positive predictions, misclassifying 14 non-entrepreneurial cases as entrepreneurial compared to only 4 missed entrepreneurial instances.

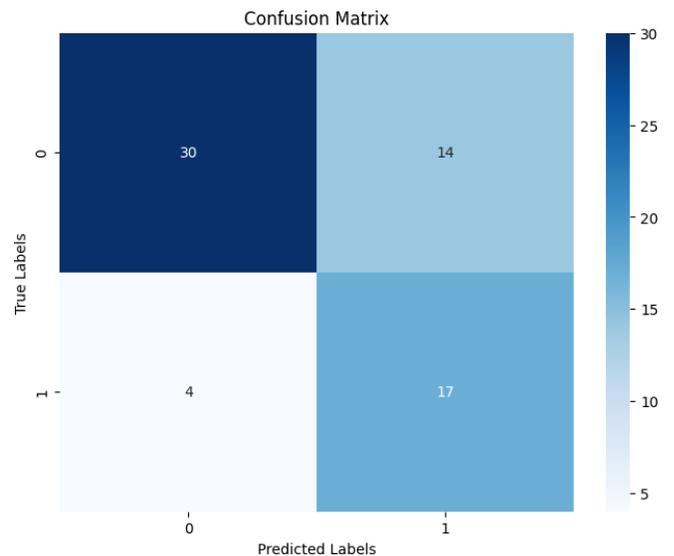

Fig. 18. Confusion Matrix for K-Nearest Neighbors

*7) Support Vector Machines:* The SVM model with an RBF kernel (C = 0.1, gamma = 'auto') produced 83.0% test accuracy. It showed strong separation capabilities with consistent recall across classes. The confusion matrix in Fig. 19 indicates that the proposed model demonstrates strong discriminative capability across both classes. Out of 65 test instances, 36 non-entrepreneurial cases and 18 entrepreneurial cases were correctly classified. Misclassification remains limited, with 8 non-entrepreneurial instances incorrectly predicted as entrepreneurial and only 3 entrepreneurial instances missed. This error distribution suggests a balanced performance. The model shows a slight bias toward correctly identifying entrepreneurial outcomes, which aligns with the aim of prioritizing empowerment in prediction tasks.

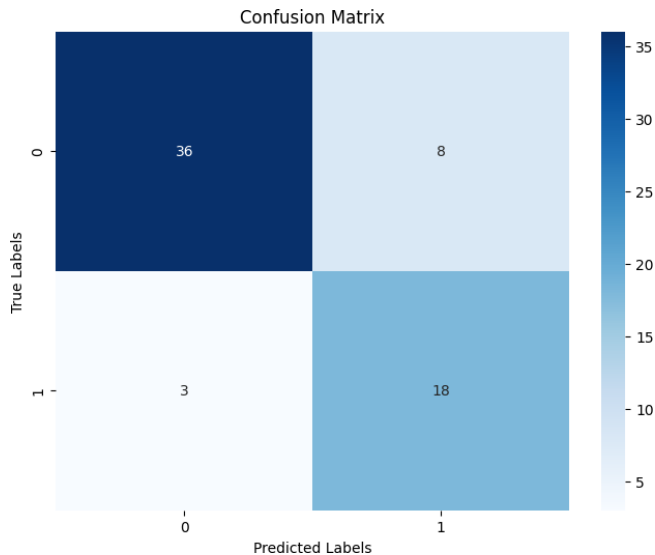

Fig. 19. Confusion Matrix for Support Vector Machines

*8) Comparative Model Summary:* Table VII-B8 summarizes the performance of all evaluated models on the test set. This evaluation reports testing Accuracy, Precision, Recall, F1-Score and AUC. Logistic Regression emerged as the top performing model achieving the highest accuracy and stable evaluation metrics.

| Model | Accuracy | Precision | Recall | F1-Score | AUC |
|---|---|---|---|---|---|
| Logistic Reg | 0.846 | 0.88 | 0.95 | 0.88 | 0.87 |
| Decision Tree | 0.615 | 0.65 | 0.81 | 0.58 | 0.67 |
| Random Forest | 0.815 | 0.85 | 0.90 | 0.82 | 0.84 |
| KNN | 0.723 | 0.77 | 0.81 | 0.73 | 0.75 |
| SVM (RBF) | 0.831 | 0.85 | 0.86 | 0.83 | 0.84 |

TABLE IX
PERFORMANCE COMPARISON OF MACHINE LEARNING MODELS

## VIII. DISCUSSION

### A. Comparison with Existing Literature

Comparing our research with existing research provides an in-depth understanding of the diverse group of teen entrepreneurs studied, and, since most of them are teenagers or young women, this adds a unique dimension to the research. It also aligns with some previous studies and the impact of social media on the macro level of economic growth and the micro level of individual women-run businesses on social media. Our research also shows the uncertainty in women and the issues they face, which policymakers can use to address them, e.g., cyberbullying, harassment, Limited access to technology and cultural norms.

### B. Implications and Significance

The implications of our research can help design initiatives to address the challenges identified. Digital education should be promoted to equip entrepreneurs with the tools they need to leverage social media effectively. Once women entrepreneurs and business leaders are empowered, they can create strategies to overcome other challenges themselves and business leaders can tailor strategies to enhance online engagement. Recognizing the role of social media is both an opportunity and a challenge. Thus, we encourage that. Collaboration should be pursued to create an environment where entrepreneurs can understand their power and the role of social media in economic growth.

### C. Limitations

While this study throws promising results, it is not without limitations. The findings are limited to teenagers and women entrepreneurs. Alongside that, the geographic areas are specific and thus do not fully capture the diverse landscape across the country. The response may be biased and based on personal experiences, thus making every respondent's thoughts and mindset about things different. Despite this, our research gives a promising understanding of the relationship between social media entrepreneurs and economic growth.

### D. Recommendations for Future Research

Future research in this area should increase sample size and include more diverse groups and geographic locations. Additionally, conducting in-depth interviews with teenagers and young women entrepreneurs may provide further insights. Governments and non-governmental organizations are encouraged to implement more effective initiatives, such as She Means Business and WomenX. The present research highlights the significance of social norms and economic conditions, which could be addressed to support the advancement of young women entrepreneurs.

## IX. CONCLUSION

This study highlights the importance of social media for young women entrepreneurs in Pakistan. While most are confident about social media's impact, some have speculations related to its potential due to concerns about traditional businesses, limited access to technology access etc. This study highlights the impacts of platforms such as Facebook, Instagram, YouTube, TikTok, Snapchat, and WhatsApp, as well as the strategies that can be implemented. However, these preferences can also be due to factors such as age and culture. This gives us more reason to market products differently across platforms and to provide more initiatives to address the unique needs and concerns that drive the economy forward. The findings show that structured analysis of social media activity successfully revealed distinct behavioural clusters, separating low-engagement viewers from highly interactive network builders. In the entrepreneurship prediction task, the models demonstrated strong reliability, with Logistic Regression and SVM emerging as the most effective and consistently accurate classifiers. Together, these results highlight the value of combining statistical evaluation with optimised machine learning techniques to uncover meaningful behavioural patterns and support robust predictive modelling.

## X. BIOGRAPHY SECTION

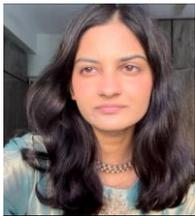

**Maryam Arif** Maryam Arif is an accomplished author with over 10 books, a social entrepreneur, and an advocate for gender equality and environmental sustainability. She was awarded the titles of 'Youngest author for 8+ books at 16 and Youngest researcher of Pakistan' for her publication [3]. Her notable works include a guide on writing and publishing in Pakistan and research on educational inequalities due to language-in-education policies. She has been recognized with multiple accolades, including being nominated for the Asian International Excellence Awards in the Female Literature category, delivering a TEDx talk and representing Pakistan at the Harvard Project for Asian and International Relations. Maryam is also a Millennium Fellow, Don Lavio Fellow, and Amal Academy Fellow. She was invited to attend the Pride of Pakistan 2024 event and has contributed to the Islamabad Women Chamber of Commerce & Industry's Trade Directory for her impactful work in business and social advocacy. Her publications also include [5] and [4].

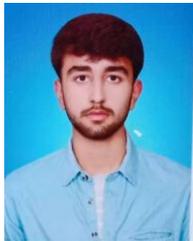

**Soban Saeed** is a researcher with a multidisciplinary background spanning biomedical engineering, electronics, embedded systems, electromedical technologies, and artificial intelligence. He has contributed to brain–computer interface (BCI) research, medical-imaging workflows, and intelligent assistive-technology design. During his research internship at the University of Alberta, he developed neurorehabilitation applications involving gait analysis, sensor-fusion algorithms, optimization methods, and clinical automation. His research interests include electronics design, embedded intelligent systems, biomedical signal processing, BCI frameworks, and Medical AI. His recent publication [5].

15digital marketing and social media usage by women entrepreneurs in pakistan. *ILMA Journal of Social Sciences & Economics*, 4(1):1–15, 2023.

[20] NS Rahayu, M Masduki, and NEE Rahayu. Women entrepreneurs and the usage of social media for business sustainability in the time of covid-19. *Research Square*, 2021. Accessed: 2025-12-10.

[21] Duaa Rehman and Urooj Qamar. Examining the influence of cultural and societal factors on networking and socializing challenges: A study of female entrepreneurs in pakistan. *Journal of Management and Research*, 11(2):53–89, 2024. Accessed: 2025-12-10.

[22] K. Ringheim. Ethical issues in social science research with special reference to sexual behaviour research. *Social Science & Medicine*, 1995.

[23] Oliver Rowntree, Kalvin Bahia, Helen Croxson, Anne Delaporte, Michael Meyer, Matthew Shanahan, and Claire Sibthorpe. The mobile gender gap report 2019. Technical report, GSMA, 2019. Accessed: 2025-12-10.

[24] N. Saif, M. W. Rana, S. Rashid, et al. Digital literacy for women entrepreneurs: Addressing the marketing knowledge gap in pakistani communities. *Journal of Social Horizons*, 2025. In press.

[25] Noor Sanauddin, Jamil Ahmad Chitrali, and Syed Owais. Public patriarchy: An analysis of women's access to education, work and politics in pakistan. *Putaj Humanities & Social Sciences*, 23(1), 2016.

[26] Philipp Schade and Monika C. Schuhmacher. Predicting entrepreneurial activity using machine learning. *Journal of Business Venturing Insights*, 19:e00357, 2023.

[27] User User. The barriers, opportunities, and policy implications for women entrepreneurs in pakistan. *Journal of Management Science Research Review*, 2025. Accessed: 2025-12-10.

[28] Dr-Syeda Waseem and Saba paracha. The study of the rising trends of digital marketing and social media usage by women entrepreneurs in pakistan. *ILMA Journal of Social Sciences Economics*, 4:1–15, 08 2023.

[29] Muhammad Zafar, Areeba Toor, and Talha Hussain. Social media as conduit for women entrepreneurs in pakistan. 01 2019.

# Survey on Social Media and Women Entrepreneurship

This survey aims to explore the various aspects of social media usage, entrepreneurial activities, challenges, and opportunities faced by teenage and young women entrepreneurs in Pakistan. (Only for women; Age 13-25) Your participation is entirely voluntary, and all responses will be kept confidential and anonymized for research purposes. Please answer each question to the best of your ability and based on your personal experiences.

* Indicates required question

Survey questionnaire

Social media and women entrepreneurship. This survey aims to explore the various aspects of social media usage, entrepreneurial activities, challenges, and opportunities faced by teenage and young women entrepreneurs in Pakistan. (Only for women; Age 13-25) Your participation is entirely voluntary, and all responses will be kept confidential and anonymized for research purposes. Please answer each question to the best of your ability and based on your personal experiences.

* Indicates required question

Your name *

Phone number

Email *

Demographic Information: What is your age? * Mark only one oval.

- 14-18
- 18-24
- 25

What is your educational background? * Mark only one oval.

- Still in school
- High school graduate College
- University Student

Social Media Usage: Mark only one oval.

- Daily
- Several times a week
- Once a week
- Rarely

How many hours per day do you spend on social media? * Mark only one oval.

- Less than 1 hour1-2
- 2-4 hours
- More than 4 hours

Are you currently engaged in any entrepreneurial activities? Mark only one oval.

- Yes
- No

How confident do you feel in utilizing social media to promote your entrepreneurial ventures? Mark only one oval.

- Very confident
- somewhat confident
- Neutral
- Not very confidentNot
- Confident at all

Are you aware of any government initiatives or programs that support youngwomen entrepreneurs in utilizing social media for business growth? Mark only one oval.

- Yes
- No

Have you used social media platforms to promote your products/services or connect with customers? Mark only one oval.

- Yes
- No

In your opinion, how has social media impacted the economic growth potential of teenage and young women entrepreneurs in Pakistan? Mark only one oval.

- Positively
- Negatively
- No impact
- Not sure

Do you believe that social media provides equal opportunities for both men and women entrepreneurs? Mark only one oval.

- Yes
- No

Have you faced any challenges or barriers in leveraging social media for your entrepreneurial pursuits? Mark only one oval.

- Yes

- No

I do not use social media for your entrepreneurial pursuits Will you use social media to sell your products in the future? Mark only one oval.

- Yes
- No
- Maybe

What factors do you believe hinder the economic growth potential of young women entrepreneurs in Pakistan in the context of social media?